\documentclass[a4paper]{jpconf}
\usepackage{graphicx}
\usepackage{bm}
\newcommand\loo{$H\!\!\parallel \!\! [100]$}
\newcommand\llo{$H\!\!\parallel \!\! [1\bar{1}0]$}
\newcommand\lll{$H\!\!\parallel \!\! [111]$}

\newcommand\GGG{$\rm Gd_3Ga_5O_{12}$}
\newcommand{\Kag}{kagom\'{e}}
\newcommand\PRB[3]{{#3} {\it Phys. Rev. B} {\bf {#1}} {#2}}
\newcommand\PPRL[3]{{#3} {\it Phys. Rev. Lett.} {\bf {#1}} {#2}}
\newcommand\JJPCM[3]{{#3} {\it J. Phys. Condens. Matter} {\bf {#1}} {#2}}

\newcommand\JJMMM[3]{{#3} {\it J. Mag. and Mag. Mater.} {\bf {#1}} {#2}}
\newcommand\APA[3]{{#3} {\it Appl. Phys. A} {\bf {#1}} {#2}}
\newcommand\PhysB[3]{{#3} {\it Physica B} {\bf {#1}} {#2}}  
\begin{document}
\title{Field induced magnetic order in the frustrated magnet Gadolinium Gallium Garnet}
\author{O~A~Petrenko$^1$, G~Balakrishnan$^1$, D~$\rm M^cK$~Paul$^1$, M~Yethiraj$^2$, G~J~McIntyre$^3$ and A~S~Wills$^4$}
\address{$^1$University of Warwick, Department of Physics, Coventry CV4~7AL, UK}
\address{$^2$Bragg Institute, ANSTO, Lucas Heights NSW 2234 Australia}
\address{$^3$Institut Laue-Langevin, 6 rue Jules Horowitz, BP 156 - 38042 Grenoble Cedex 9, France}
\address{$^4$Department of Chemistry, University College London, 20 Gordon Street, London WC1H~0AJ, UK}
\ead{O.Petrenko@warwick.ac.uk}
\begin{abstract}
\GGG, (GGG), has an extraordinary magnetic phase diagram, where no long range order is found down to 25~mK despite $\Theta_{CW}\!\!\approx $2~K.
However, long range order is induced by an applied field of around 1~T.
Motivated by recent theoretical developments and the experimental results for a closely related hyperkagome system, we have performed neutron diffraction measurements on a single crystal sample of GGG  in an applied magnetic field.
The measurements reveal that the $H-T$ phase diagram of GGG is much more complicated than previously assumed.
The  application  of an external  field at low $T$ results in an intensity change for most of the magnetic peaks which can be divided into three distinct sets: ferromagnetic, commensurate antiferromagnetic, and  incommensurate antiferromagnetic.
The ferromagnetic peaks (e.g. (112), (440) and (220)) have intensities that increase with the field and saturate at high field.
The antiferromagnetic reflections have intensities that grow in low fields, reach a maximum at an intermediate field (apart from the $(002)$ peak which shows two local maxima) and then decrease and disappear above  2~T.
These AFM peaks appear, disappear and reach maxima in {\it different} fields.
We conclude that the competition between magnetic interactions and alternative ground states prevents  GGG from ordering in zero field. It is, however, on the verge of ordering and an applied magnetic field can be used to crystallise ordered components. The range of ferromagnetic (FM) and antiferromagnetic (AFM) propagation vectors found reflects the complex frustration in GGG.
\end{abstract}

\section{Introduction}
\begin{figure}[t]
\includegraphics[width=16pc]{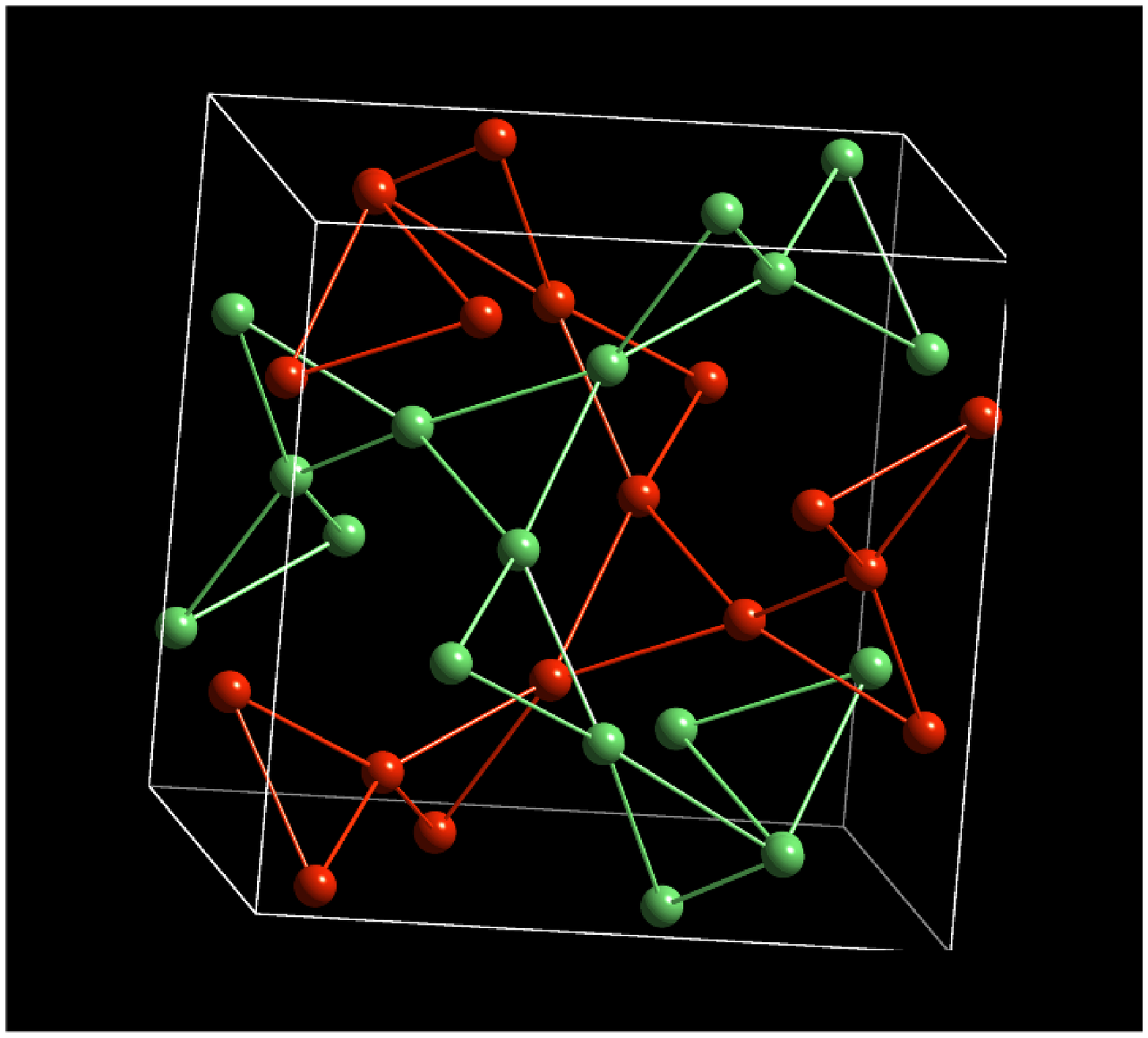}\hspace{2pc}
\begin{minipage}[b]{19pc}\caption{\label{Fig_structure}Positions of the magnetic Gd ions in a garnet structure. There are 24 magnetic ions per unit cell, they are divided into two interpenetrating sublattices.}
\end{minipage}
\end{figure}

Gadolinium Gallium Garnet, \GGG, (GGG), occupies a unique position among the geometrically frustrated magnetic systems.
It has been a subject of intense research for the past three decades, as in terms of the geometry it lies somewhere between the well-studied case of {\it stacked triangular} lattices and lattices with a much more severe degree of frustration, such as the \Kag\ lattice antiferromagnets and the pyrochlores.
In GGG the magnetic Gd ions are located on two interpenetrating corner-sharing triangular sublattices, as shown in Fig.~\ref{Fig_structure}.
It has been known for more than twenty years that GGG has an extraordinary low-temperature phase diagram~\cite{GGG_Hov_80,GGG_Schiffer_94}.
No long range magnetic order (LRO) has been found in GGG down to 25~mK, LRO can be induced only by an applied  magnetic field above  $\sim1$~T.
The magnetic phase diagram depends on the external field orientation, as the critical temperatures and critical fields for \loo\ and \lll\ differ by 30-40\%~\cite{GGG_Hov_80}.
Another peculiar feature of the $H-T$ phase diagram of GGG it that the phase boundary between the LRO and spin-liquid phases has a distinct minimum at $T\approx 0.18$~K, analogous to the minimum in the melting curve of $^4$He~\cite{GGG_Tsui_99}.
In this paper we report neutron diffraction measurements on a single crystal sample of GGG performed in an applied magnetic field at low temperatures.
These measurements reveal that the $H-T$ phase diagram of GGG has a much more complicated nature than previously assumed.

The interest in \GGG\ has been renewed recently by several developments.
Firstly, a new mean-field approach to the theoretical studies of GGG has been developed for zero magnetic field~\cite{GGG_Taras1,GGG_Taras2}.
The theory suggests that at low temperature GGG is on the verge of achieving true LRO~\cite{GGG_Taras3}. 
Secondly, the latest experimental and theoretical studies of a closely related ``hyperkagome" system, Na$_4$Ir$_3$O$_8$, have considered the role of classical and quantum fluctuations in lifting the macroscopic degeneracy of the apparent spin-liquid state~\cite{HyperKagome}.
The similarity between the lattice structures of Na$_4$Ir$_3$O$_8$ and \GGG\ is obvious and further comparison of their magnetic properties would be valuable.
Finally, the reasons for GGG being a very efficient material for low temperature magnetic refrigeration have been reconsidered in view of the presence of a macroscopic number of soft modes in frustrated magnets below the saturation field~\cite{cooling}.  

\section{Experimental details}
Polycrystalline GGG was synthesised from stoichiometric quantities of Gd$_2$O$_3$ and Ga$_2$O$_3$ by a solid diffusion reaction at $T=1400^\circ$C  for 12 hours with intermediate regrinding.
The procedure of regrinding was repeated until an X-ray diffraction study showed no impurity phases in the final product.
In order to minimise neutron absorption we have prepared a sample with 99.98\% of the non-absorbing isotope, $^{160}$Gd; this isotope was supplied by Oak Ridge National Laboratory.
This polycrystalline sample was used previously for neutron diffraction measurements in zero field~\cite{GGG_powder_97_98} and in an applied magnetic field~\cite{GGG_powder_99}.
A single crystal sample of GGG was then grown by the floating zone method using a two mirror infra-red image furnace.
The low-temperature specific heat measurements performed on the single crystal sample~\cite{C_lowT_Canada} have shown results that are very similar to the results reported by Schiffer {\it et al.}~\cite{GGG_Schiffer_95}, which implies an absence of a significant amount of impurity. 

The single crystal neutron scattering measurements were carried out using the D10 instrument at Institut Laue-Langevin, France.
An $80 \times 80$~mm$^2$ two-dimensional microstrip detector was used in the diffraction configuration.
Typical intensities of the main nuclear Bragg peaks for the incident wavelength neutrons of $\lambda = 2.36$~\AA\ were several hundred counts per second.
We have measured the magnetic diffraction patterns at temperatures between 50~mK and 0.7~K in fields of up to 2.5~T. 
The horizontal scattering plane contained the $(hhl)$ reflections so that an external magnetic field provided by a vertical cryomagnet was applied along the [$1\bar{1}0$] direction.
The results of the measurements in the $(hk0)$ scattering plane for a field applied along the [100] direction performed on the same sample at Berlin Neutron Scattering Center, Germany have been reported previously~\cite{GGG_Xtal_02}.

Apart from describing our new single crystal results, we have also reanalysed previously reported neutron powder diffraction data~\cite{GGG_powder_99} and indexed the field-induced incommensurate magnetic Bragg peaks. 

\section{Results and Discussion}
Prior to the application of a magnetic field, we have attempted to find the relatively sharp low-intensity magnetic Bragg peaks, which have been detected in a powder diffraction pattern at temperatures below 140~mK at incommensurate positions, such as $Q_1=0.64$~\AA$^{-1}$ and $Q_2=0.85$~\AA$^{-1}$ and several others.
Recent theoretical results~\cite{GGG_Taras1,GGG_Taras2,GGG_Taras3} advocated a scenario that in the zero field regime GGG upon cooling is very close to developing conventional long range magnetic order, which corresponds to the appearance of the sharp but not resolution limited diffraction peaks.
These peaks give a correlation length of approximately 100~\AA, suggesting extended short range order (over 8 cubic unit cells of 12.3~\AA\ each). 
Our attempts to find peaks of the $(hhl)$ type with a scattering vector length equal to $Q_1$ or $Q_2$ in single crystal experiment were unsuccessful.
\begin{figure}[tb]
\begin{minipage}{18pc}
\includegraphics[width=17.5pc]{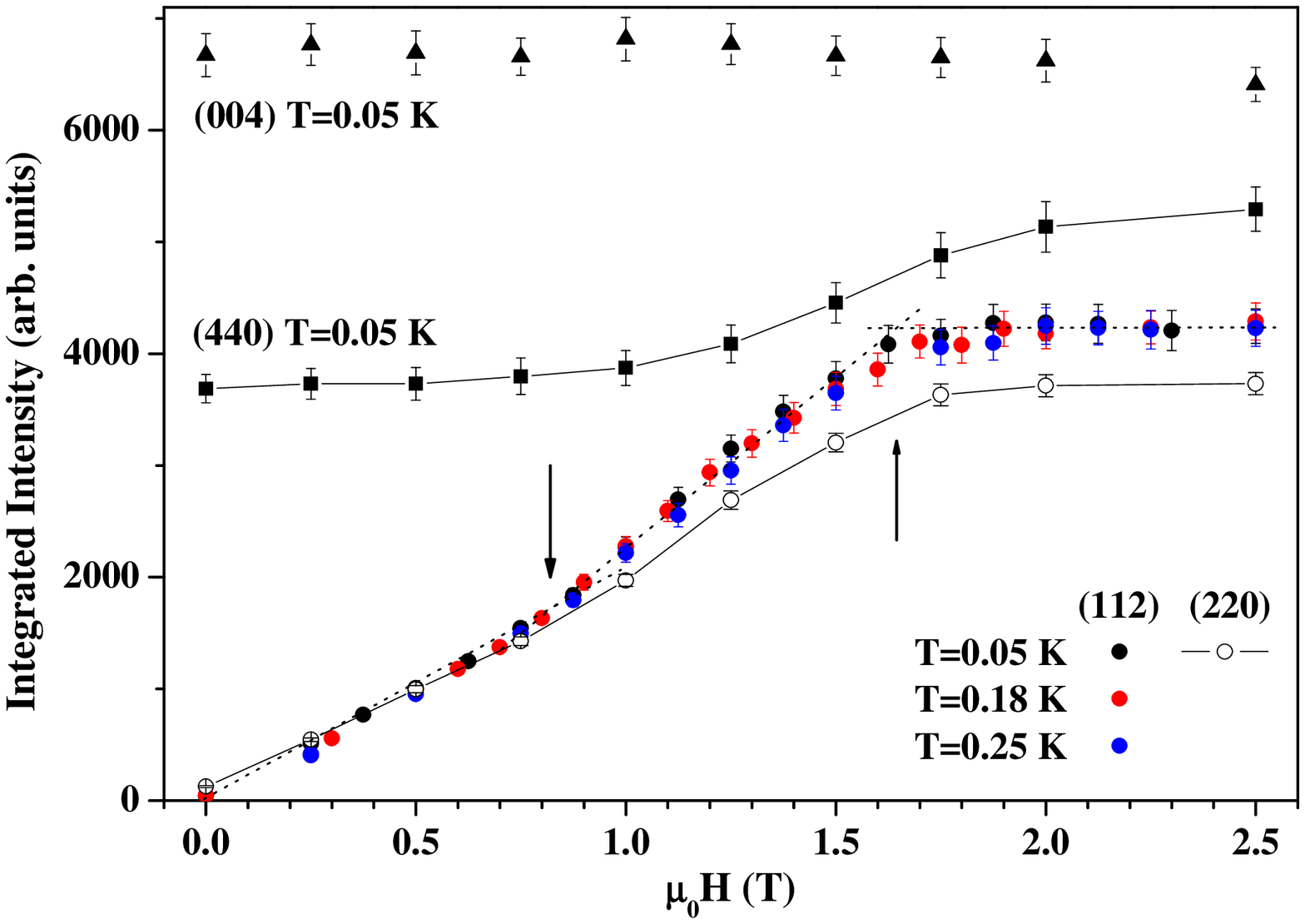}
\caption{\label{Fig_label2} Field dependence of the integrated intensity of the three peaks with a strong FM component, $(112)$, $(220)$ and $(440)$ as measured in a GGG single crystal for \llo.
                           The arrows indicate the fields of 0.82~T and 1.65~T where the peaks exhibit a change in a rate with which the intensity rises with field.
                           The intensity of the field-independent nuclear peak $(004)$ is also shown for reference.}
\end{minipage}\hspace{1.5pc}%
\begin{minipage}{18pc}
\includegraphics[width=17.5pc]{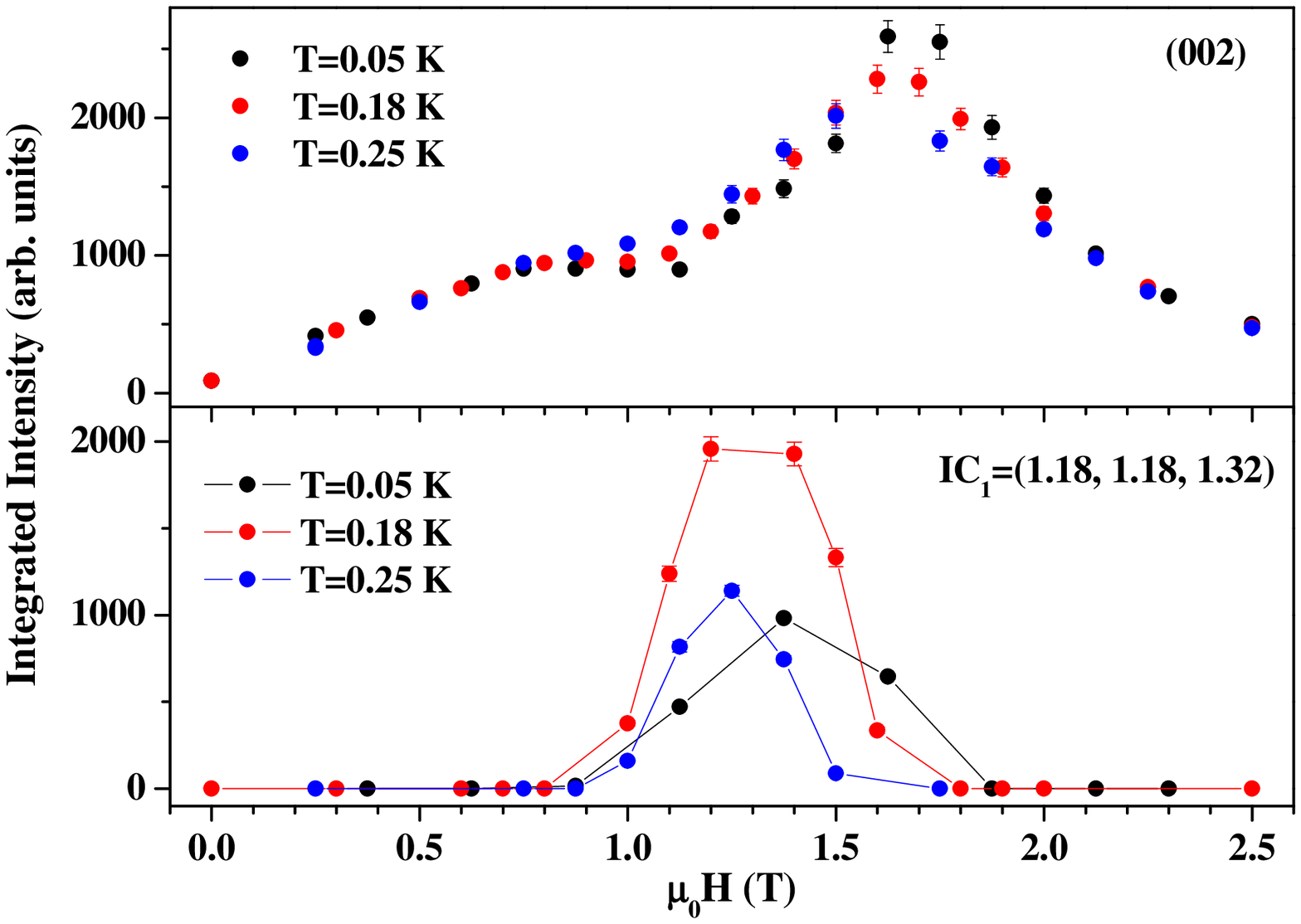}
\caption{\label{Fig_label3}Field dependence of the intensity of the two AFM peaks measured at different temperatures in a GGG single crystal for \llo.
                           Top panel -- $(002)$ peak, bottom panel -- an incommensurate peak nominally observed at IC$_1=(1.18, 1.18, 1.32)$ slightly off-centre of the 2D position sensitive detector.
                           The peak is indexed as $(1.35, 1.05, 1.37)$ when off-centring is taken into account.}
\end{minipage} 
\end{figure}

The  application  of an external  magnetic  field at low temperature results in an intensity change for most of the peaks, which could be divided into the three main sets, ferromagnetic, commensurate antiferromagnetic and incommensurate antiferromagnetic.
The  ferromagnetic peaks have intensities that increase with the field and saturate at some higher field, while the antiferromagnetic peaks have intensity  that grows in low  fields, reaching a maximum at some intermediate field then decreases and (almost) disappears at fields  above  2~T. 

Fig.~\ref{Fig_label2} shows the field dependence of the integrated intensity, $I$, of the three FM peaks $(112)$, $(220)$ and $(440)$ as well as of one purely nuclear peak $(004)$ measured at the base temperature of 50~mK.
One immediate observation from this figure is that there are two fields at which the gradient of the $I(H)$ curves exhibit a change: a pronounced flattening at 1.65~T presumably associated with magnetisation saturation and another change at half this field where the gradient increases by approximately 50\%.
The changes are most noticeable in the field dependence of the $(112)$ peak, as the experimental point density there is the highest, but the same tendency is clearly visible for the $(220)$ and $(440)$ peaks as well.
The data in Fig.~\ref{Fig_label2} also indicate that there is no significant temperature dependence of the intensity of the $(112)$ peak for 50~mK~$< T <$~250~mK.
Nuclear contributions to the $(112)$ and $(220)$ reflections are allowed by symmetry, but this component is small compared to the FM field-induced intensity.
\begin{table}[tb]
\caption{\label{table1}A description of the various k-vectors present in GGG powder profiles~\cite{GGG_powder_99} at $T=0.08$~K in different fields determined by reverse-Monte Carlo indexing~\cite{Sarah,B-Mn}.} 
\begin{center}
\lineup
\begin{tabular}{*{4}{l}}
\br                              
Field, T&$k_1=(000)$&$k_2=(001)$&$k_3=(0,0,0.724)$\cr 
\mr
\0 1.0&\0\0 $\bullet$ &\0\0 $\bullet$ &\0\0\0\0 $\bullet$ \cr
\0 1.6&\0\0 $\bullet$ &\0\0 $\bullet$ &\0\0\0\0 $-$\cr 
\0 3.3&\0\0 $\bullet$ &\0\0 $-$&\0\0\0\0 $-$\cr 
\br
\end{tabular}
\end{center}
\end{table}

\begin{figure}[b]
\begin{minipage}{18pc}
\includegraphics[width=17.5pc]{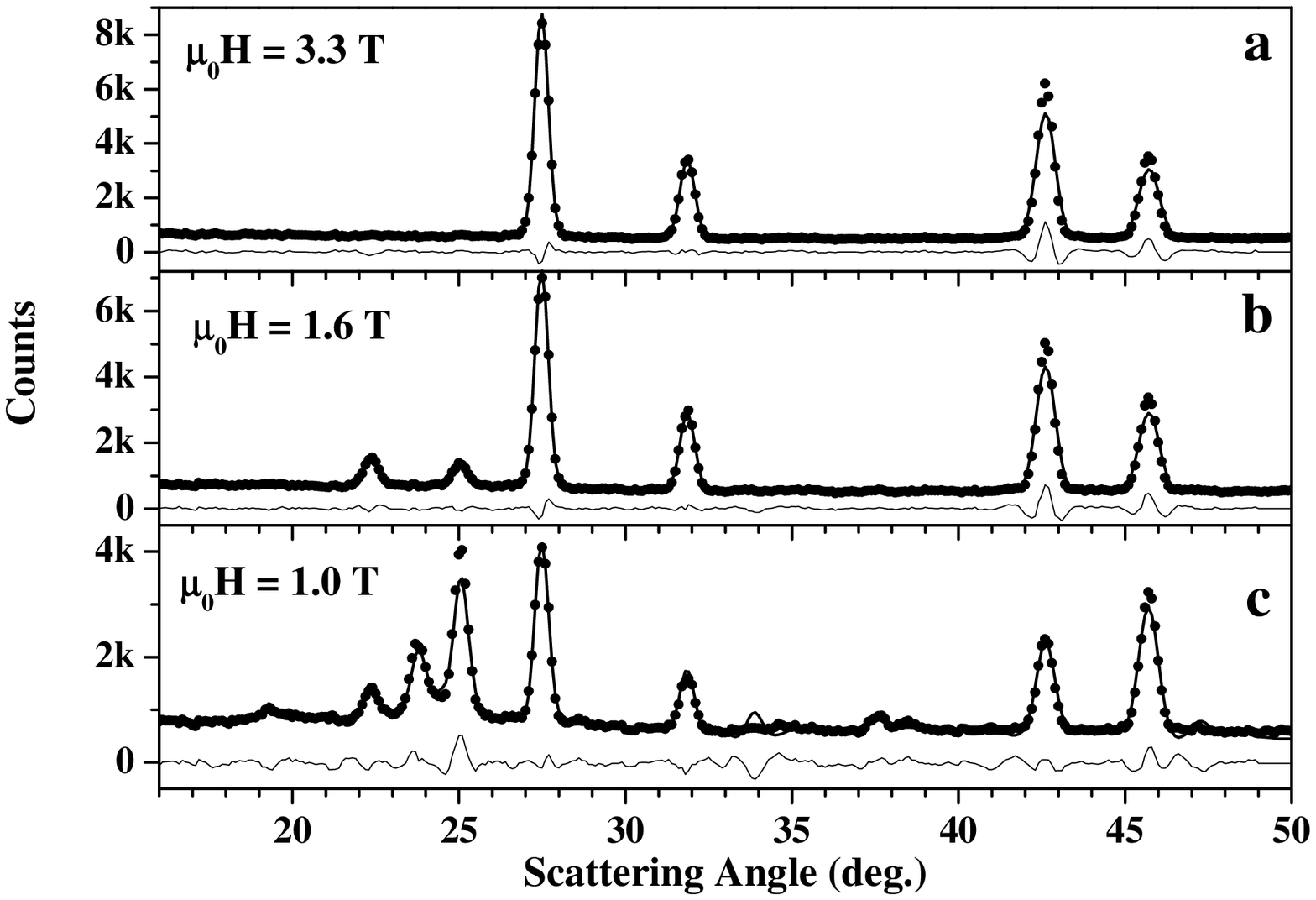}
\caption{\label{Fig_label4} Observed (symbols), calculated (full lines) and difference neutron powder diffraction profiles for GGG~\cite{GGG_powder_99} in an applied field of (a) 3.3~T, (b) 1.6~T and (c) 1.0~T at $T=80$~mK. \\ }
\end{minipage}\hspace{1.5pc}
\begin{minipage}{18pc}
\includegraphics[width=17.5pc]{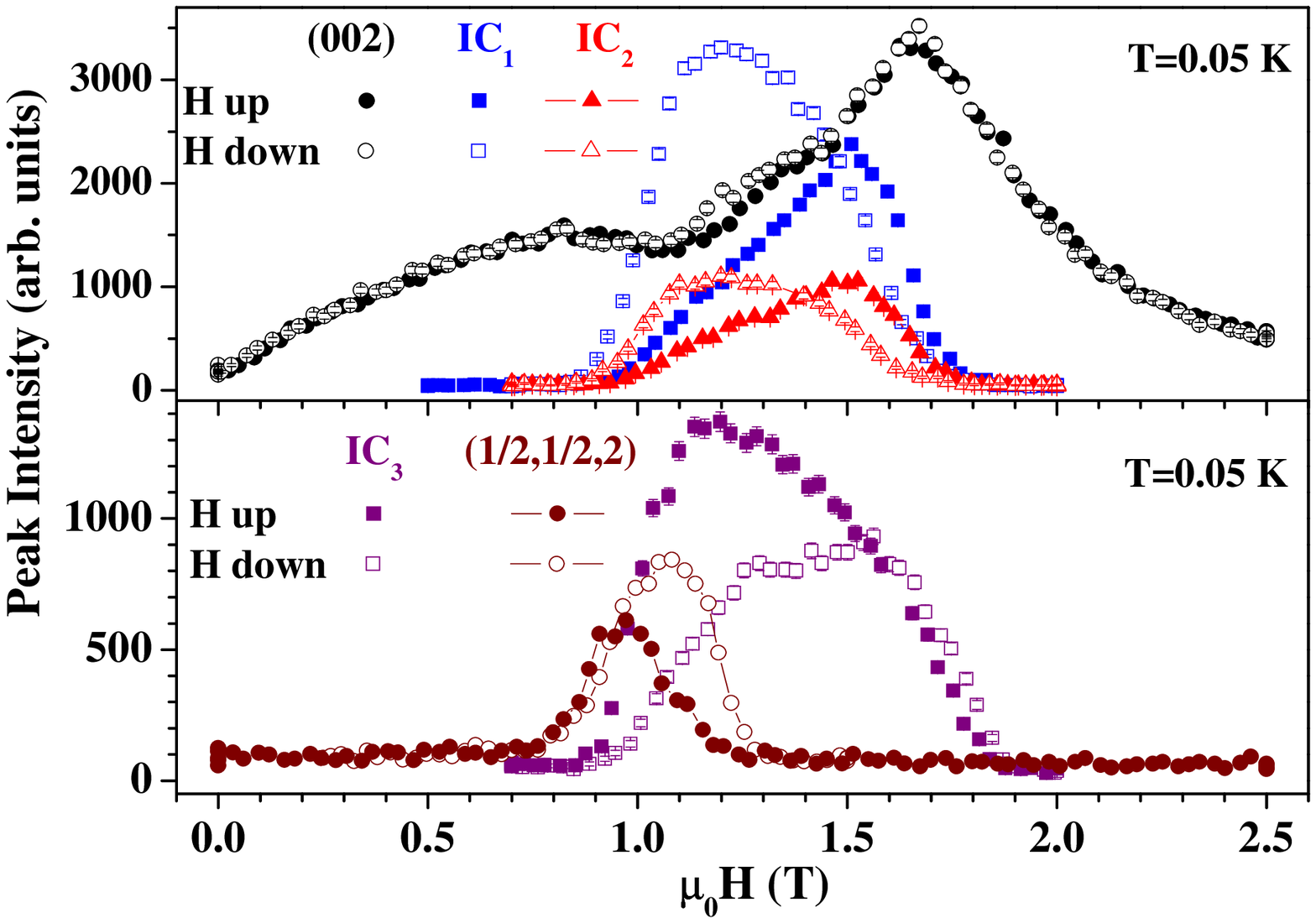}
\caption{\label{Fig_label5}Field dependence of the AFM peaks intensity, as measured by ramping an external field \llo\ up/down (solid/open symbols) with a rate of $\approx 0.1$~T/min.
                                                IC$_{1,2,3}$ denote the incommensurate peaks indexed in the main text.}
\end{minipage} 
\end{figure}
Fig.~\ref{Fig_label3} shows the evolution of the intensity of two AFM peaks (commensurate and incommensurate) with an applied magnetic field.
These peaks are purely magnetic as the translational symmetry of the crystal structure prevents any nuclear contribution.
One of the peaks, $(002)$, clearly shows two local maxima  at the base temperature of 50~mK.
The positions of these maxima (0.8~T and 1.65~T) match well the fields where the $I(H)$ curves for the FM peaks show a change in gradient.
Another  peak, IC$_1$, is  incommensurate, indexed as $(1.35, 1.05, 1.37)$, but is nominally observed at a point $(1.18, 1.18, 1.32)$ slightly off-centre on the 2D position sensitive detector.
Unlike the FM peaks, the IC$_1$ is sensitive to small temperature variations (see Fig.~\ref{Fig_label3} bottom panel).
Apart from the AFM peaks shown in Fig.~\ref{Fig_label3}, we have observed similar behaviour for the half-integer peak $(\frac{1}{2} \frac{1}{2} 2)$ and the incommensurate peaks IC$_2=(-1.53, -1.53, 0.51)$, IC$_3=(0.59, 0.59, 1.95)$ and IC$_4=(1.46, 1.46, -0.51)$  indexed as $(-1.48, -1.56, 0.51)$, $(0.52, 0.68, 2.04)$ and $(1.41, 1.60, -0.51)$ respectively when off-centring is taken into account.
The relatively small number of the magnetic reflections observed (largely limited to an $(hhl)$ scattering plane) does not allow for a full structure refinement.
However, even if a larger number of peaks was available, the magnetic structure determination would not be straightforward, as the AFM peaks appear and disappear in slightly different magnetic fields.
This is well demonstrated in Figure~\ref{Fig_label4} where the incommensurate and then commensurate antiferromagnetic peaks disappear upon increasing the field.
The refined propagation vectors and their presence in the different field regimes are shown in Table~\ref{table1}.
The peaks are not resolution limited, indicating that the field-induced magnetic order in GGG is short range in nature.

The data in Figs.~\ref{Fig_label2} and~\ref{Fig_label3} were collected by fixing an external field and performing $\omega$-scans for each reflection.
This method ensures an accurate measurement of the intensity and the position of the Bragg peaks.
Alternatively, a much more rapid data collection is achieved by constantly ramping an external magnetic field and simply counting the number of scattered neutrons per second.
Although this method is less accurate in terms of measuring absolute peak intensity (particularly if the peaks are changing their positions or widths with field), it allows the study of any sample history dependence and  observation of possible metastable states.
The results of such an approach are summarised in Fig.~\ref{Fig_label4}.
As can be seen from this Figure, the incommensurate peaks are very sensitive to the sample's field and temperature history and demonstrate significant hysteresis.
Less pronounced hysteresis is also observed for a $(002)$ peak between 1.1~T and 1.4~T.
Remarkably, the $(\frac{1}{2} \frac{1}{2} 2)$ seems to exist only in the vicinity of $(1\pm 0.2)$~T, while the incommensurate peaks have non-zero intensity over a much wider region between 0.8~T and 1.8~T.

\begin{figure}[tb]
\begin{minipage}{18pc}
\includegraphics[width=17.5pc]{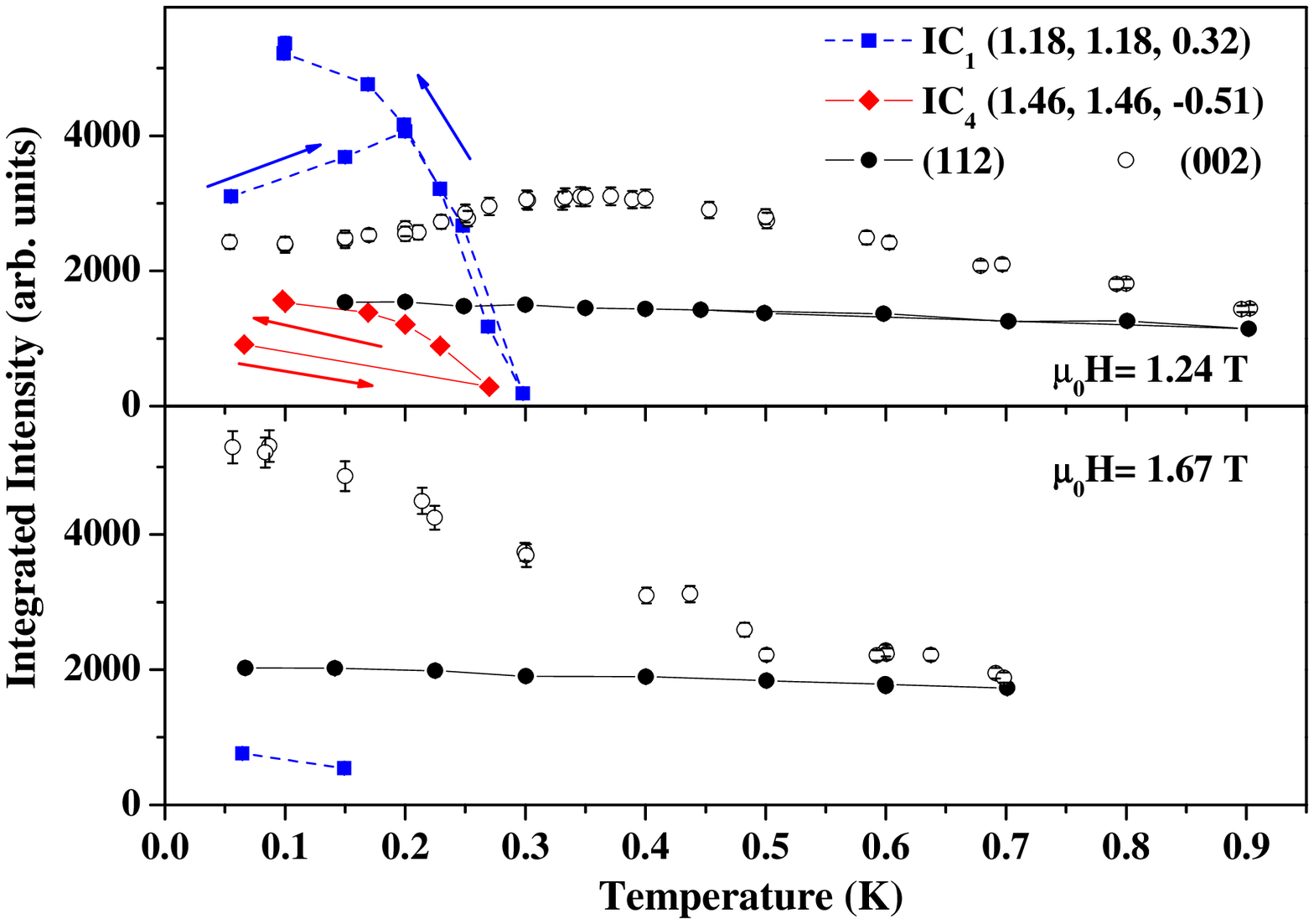}
\caption{\label{Fig_label6}$T$ dependence of the peaks intensity for \llo\ at 1.24~T and 1.67~T (top and bottom panels).
                                               The arrow on the upper panel indicate samples cooling or warming.
                                                For the lower panel the data were collected in a field warming regime.
                                                The notations IC$_{1,4}$ correspond to the incommensurate peaks, their indices are given in the main text.}
\end{minipage}\hspace{1.5pc}
\begin{minipage}{18pc}
\includegraphics[width=17.5pc]{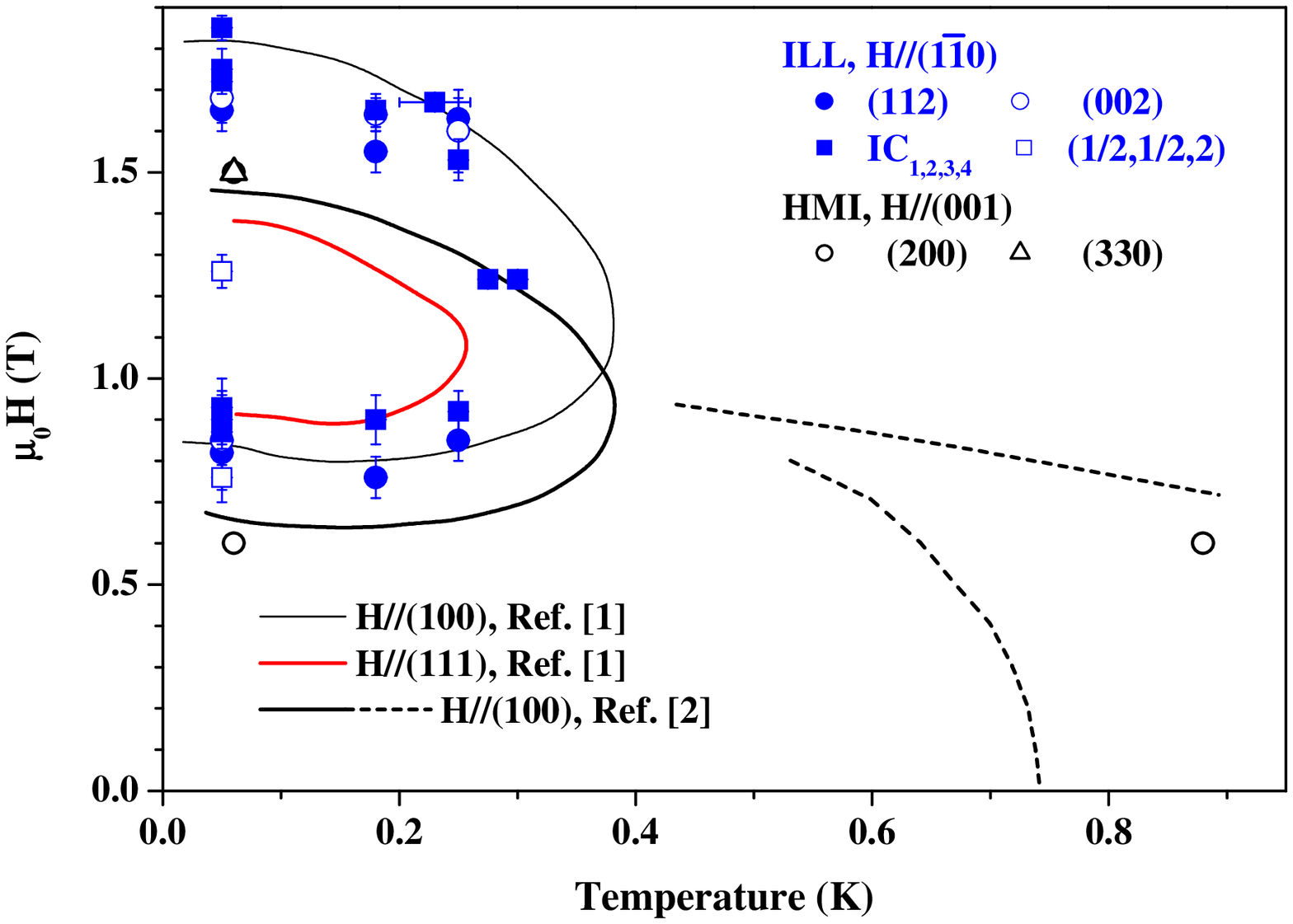}
\caption{\label{Fig_label7} A comparison of the magnetic phase diagram of GGG determined by bulk properties measurements (lines)~\cite{GGG_Hov_80,GGG_Schiffer_94} and neutron diffraction results (symbols).
                                                 Solid and dashed lines correspond to the suggested regions of LRO and SRO.
                                                 Apart from the new ILL results, neutron data from the previous HMI experiment~\cite{GGG_Xtal_02} are also shown.}
\end{minipage} 
\end{figure}

Having seen which peaks appear at base temperature in an applied magnetic field, it is equally important to follow the temperature dependence of these peaks.
Fig.~\ref{Fig_label5} shows the temperature dependence of the intensity of the selected FM and AFM peaks in a field of 1.24~T and 1.67~T.
A large difference between the results for field-cooling and field-warming regimes for the incommensurate peaks again points to a metastable nature of the field-induced state.
The instability and history-dependence were seen in the Monte Carlo simulations of GGG~\cite{GGG_MC_01}, although doubts were raised whether the energy of dipole-dipole interactions was  calculated properly~\cite{GGG_Taras1,GGG_Taras2}. 

Fig.~\ref{Fig_label6} shows the magnetic phase diagram of GGG, where our neutron scattering results are compared to the bulk properties measurements~\cite{GGG_Hov_80,GGG_Schiffer_94}.

\section{Conclusions}
To summarise, GGG is an intriguing magnetically frustrated system where it is difficult to project out a single ground state in an applied magnetic field.

\end{document}